\documentclass[sigconf]{acmart}
\usepackage{listings}
\graphicspath{{figures/}}
\DeclareGraphicsExtensions{.pdf,.jpeg,.png}
\usepackage{url}
\usepackage{color}
\usepackage{amsmath}
\usepackage{flushend}
\usepackage{multirow}

\newcommand{\code}[1]{\texttt{#1}}

\newcommand{\figref}[1]{Figure~\ref{#1}}

\newcommand{\secref}[1]{Section~\ref{#1}}
\newcommand{\tblref}[1]{Table~\ref{#1}}
\newcommand{\smalltitle}[1]{{\smallskip \noindent \bf  {#1}.\ }}

\usepackage{listings}

\newtoggle{longversion}
\toggletrue{longversion}

\newtoggle{patchtype}
\togglefalse{patchtype}

\begin{document}

\copyrightyear{2018} 
\acmYear{2018} 
\setcopyright{acmcopyright}
\acmConference[ICSE '18]{ICSE '18: 40th International Conference on
Software Engineering }{May 27-June 3, 2018}{Gothenburg, Sweden}
\acmBooktitle{ICSE '18: ICSE '18: 40th International Conference on Software
Engineering , May 27-June 3, 2018, Gothenburg, Sweden}
\acmPrice{15.00}
\acmDOI{10.1145/3180155.3180182}
\acmISBN{978-1-4503-5638-1/18/05}

\title{Identifying Patch Correctness in Test-Based Program Repair}

\author{{Yingfei Xiong, Xinyuan Liu, Muhan Zeng, Lu Zhang, Gang Huang}}
\thanks{The authors acknowledge the anonymous reviewers for the constructive comments and revision suggestions. This work is supported by the National Key Research and Development Program under Grant No. 2016YFB1000105, and National Natural Science Foundation of China under Grant No. 61725201, 61529201, 61725201, 61672045. Lu Zhang is the corresponding author. Xinyuan Liu and Muhan Zeng are equal contributors to the paper and their names are sorted alphabetically.}
\affiliation{Key Laboratory of High Confidence Software Technologies (Peking University), MoE\\
Institute of Software, EECS, Peking University, Beijing, 100871, China\\
}
\email{{xiongyf,liuxinyuan,mhzeng,zhanglucs,hg}@pku.edu.cn}


\begin{abstract}
  Test-based automatic program repair has attracted a lot of attention in recent years. 
  However, the test suites in practice are often too weak to guarantee correctness
and existing approaches often generate a large number of incorrect patches. 
  
  To reduce the number of incorrect patches generated, we propose a novel approach that heuristically determines the correctness of the generated patches. The core idea is to exploit the behavior similarity of test case executions. The passing tests on original and patched programs are likely to behave similarly while the failing tests on original and patched programs are likely to behave differently. Also, if two tests exhibit similar runtime behavior, the two tests are likely to have the same test results. Based on these observations, we generate new test inputs to enhance the test suites and use their behavior similarity to determine patch correctness.

  Our approach is evaluated on a dataset consisting of 139 patches generated from existing program repair systems including jGenProg, Nopol, jKali, ACS and HDRepair. Our approach successfully prevented 56.3\% of the incorrect patches to be generated, without blocking any correct patches.




\end{abstract}


\maketitle

%

\section{Introduction}


In the past decades, a large number of automated program repair
approaches~\cite{GenProgTSE, xiong-icse17, Gao2015Fixing, PAR, staged, Xuan2016History, gupta2017deepfix, Long2016, Gao, Mechtaev, SemFix, DirectFix, xiong2015range, liu2018sofix} have been proposed, and many of them fall into the
category of test-based program repair. In test-based program repair,
the repair tool takes a faulty program and a test suite
including at least one failing test that reveals the fault as input and then generates a
patch that makes all tests pass.
However, test suites in real world projects are often weak~\cite{qi15}, and  a
patched program passing all the tests may still be faulty. We call a patch \emph{plausible} if the patched version
passes all tests in the test suite, and we consider a patch \emph{correct} if it fixes and only fixes the bug. As studied by Long et
al.~\cite{long2016analysis}, the test suites in real world systems are
usually weak such that most of the plausible patches are incorrect, making it
difficult for a test-based program repair system to ensure the
correctness of the patches. As existing
studies~\cite{qi15,smith2015cure,martinez:hal-01321615} show, multiple
automatic program repair systems produce much more incorrect patches
than correct patches on real world defects, leading to low \emph{precision} in their generated patches. 

The low precision of existing program repair systems significantly affects the usability of these systems. 
Since test suites cannot guarantee the correctness of the patches,
developers have to manually verify patches. When the precision of a
program repair system is low, the developer has to verify a lot of
incorrect patches, and it is not clear whether such a verification 
process is more costly than directly repairing the defect by the
developers. 
An existing study~\cite{Tao:2014:AGP:2635868.2635873} also
shows that, when developers are provided with low-quality patches,
their performance will drop compared to the situation where no patch
is provided. As a result, we believe it is critical to improve the
precision of program repair systems, even at the risk of losing
some correct patches. 


Since a weak test suite is not enough to filter out the incorrect patches produced by program repair systems, a direct idea is to enhance the test suite. 
Indeed, existing studies~\cite{Xin17,Yang17} have attempted to generate new test cases to identify incorrect patches. However, while test inputs can be generated, test oracles cannot be automatically generated in general, known as the oracle problem~~\cite{oracle_survey, pezze2015automated}. As a result, existing approaches either require human to determine test results~\cite{Xin17}, which is too expensive in many scenarios, or rely on inherent oracles such as crash-free~\cite{Yang17}, which can only identify certain types of incorrect patches that violate such oracles. 



Our goal is to classify patches heuristically without knowing the full oracle. 
Given a set of plausible patches, we try to determine whether each patch is likely to be correct or incorrect, and reject the patches that are likely to be incorrect.
Our approach is based on two key observations.
\begin{itemize}
\item {\bf PATCH-SIM}. After a correct patch is applied, a passing test usually behaves similarly as before, while a failing test usually behaves differently. 
\item {\bf TEST-SIM}. When two tests have similar executions, they are
  likely to have the same test results, i.e., both triggering the same
  fault or both are normal executions.
\end{itemize}
PATCH-SIM allows us to test patches without oracles, i.e., we
run the tests before and after patching the system and check the
degree of behavior change. As our evaluation will show later, PATCH-SIM alone already identify a large set of incorrect patches. However, we can only utilize the original
tests but not the newly generated test inputs as we
do not know whether they pass or fail. TEST-SIM \iftoggle{longversion}{
 allows us to use newly generated test inputs: if a generated test input behaves similarly to an original passing test, it is likely to be a passing test. Similarly, if it behaves similarly to a failing test, it is likely to be a failing test. 
 }{complements PATCH-SIM by determining the test results of newly generated test inputs.}
\iftoggle{longversion}{
Please note that TEST-SIM does not solve the oracle problem: a test classified as failing is unlikely to expose new faults. Nevertheless, TEST-SIM is useful in complementing PATCH-SIM for patch classification.}{}

Based on these two key observations, our approach consists of the following
steps. First, we generate a set of new test inputs. Second, we
classify the newly generated test inputs as passing or failing tests by
comparing them with existing test inputs. Third, we determine the
correctness of the patch by comparing the executions before and after
the patch for each test, including both the original and the generated
tests.

We have realized our approach by designing concrete formulas to
compare executions, and evaluated our approach on a dataset of 139
patches generated from previous program repair systems including
jGenProg~\cite{martinez:hal-01321615},
Nopol~\cite{martinez:hal-01321615}, jKali~\cite{martinez:hal-01321615},
HDRepair~\cite{Xuan2016History}, and ACS~\cite{xiong-icse17}. Our
approach successfully filtered out 56.3\% of the incorrect patches 
without losing any of the correct patches.
The results indicate
that our approach increases the precision of program repair
approaches with limited negative impact on the recall. 


In summary, the paper makes the following main contributions.
\begin{itemize}
\item We propose two heuristics, {PATCH-SIM} and
  {TEST-SIM}, which provide indicators for patch correctness.
\item We design a concrete approach that automatically classifies patches
  based on the two heuristics.
\item We have evaluated the approach on a large set of patches, and
  the results indicate the usefulness of our approach.
\end{itemize}


The rest of the paper is organized as follows. \secref{sec:related} first discusses related work. \secref{sec:motivation} motivates our approach with examples, and \secref{sec:approach} introduces our approach in details. \secref{sec:implementation} introduces our implementation. \secref{sec:eval} describes our evaluation on the dataset of 139 patches. \secref{sec:threats} 
discusses the threats to validity.
Finally, \secref{sec:conclusion} concludes the paper.

\section{Related Work \label{sec:related}}
\vspace{-2mm}
\smalltitle{Test-based Program Repair}
Test-based program repair is often treated as a
search problem by defining a search space of patches, usually through
a set of predefined repair templates, where the goal is to locate
correct patches in the search space. 
Typical
ways to locate a patch include the follows.
\begin{itemize}
\item \emph{Search Algorithms.} Some approaches use meta-heuristic~\cite{GenProgTSE,staged} or random~\cite{RSRepair} search
  to locate a patch.
\item \emph{Statistics.} Some approaches build a statistical model to
  select the patches that are likely to
  fix the defects based on various information sources, such as
  existing patches~\cite{PAR,Long2016,Xuan2016History} and existing source code~\cite{xiong-icse17}. 
\item \emph{Constraint Solving.} Some approaches~\cite{SemFix,DirectFix,Mechtaev,dqlose,Singh:2013:AFG:2491956.2462195,S3}
  convert the search problem to a satisfiability or optimization
  problem and use constraint solvers to locate a patch.
\end{itemize}

While the concrete methods for generating patches are different, weak test suites problem still remains as a challenge to test-based program repair and may lead to incorrect patches generated. As our evaluation has shown, our approach can
effectively augment these existing approaches to raise their
precisions.

\smalltitle{Patch Classification}
Facing the challenge of weak test suites, several researchers also
propose approaches for determining the correctness of patches. Some researchers seek for deterministic approaches. Xin and Reiss~\cite{Xin17} assume the existence of a perfect oracle (usually manual) to classify test results and generate new test inputs to identify oracle violations. Yang et
al.~\cite{Yang17} generate test inputs and monitor the violation of inherent oracles, including crashes and memory-safety problems. Compared with them, our approach does not need a perfect oracle and can potentially identify incorrect patches that do not violate inherent oracles, but has the risk of misclassifying correct patches.

Other approaches also use heuristic means to classify patches.
Tan et
al.~\cite{tananti} propose anti-patterns to capture typical incorrect
patches that fall into specific static structures. Our approach mainly relies on dynamic information, and as the evaluation will show, the two approaches can potentially be combined.
Yu et al.~\cite{Yu17} study the approach that filters patches by
minimizing the behavioral impact on the generated tests, and find
that this approach cannot increase the
precision of existing program repair approaches. Compared with their
approach, our approach classifies the generated tests and puts
different behavioral requirements on different classes. Finally, Weimer et al.~\cite{Weimer2016Trusted} highlight possible directions
 in identifying the correctness of patches.

\smalltitle{Patch Ranking}
Many repair approaches use an internal ranking component that
ranks the patches by their probability of being correct. 
Patch ranking is a very related but different problem from patch classification. On the one hand, we can convert a patch classification problem into a patch ranking problem by setting a proper threshold to distinguish correct and incorrect patches. On the other hand, a perfect patch ranking method does not necessarily lead to a perfect patch classification method, as the threshold can be different from defect to defect. 

There are three main categories of patch ranking techniques. The first ranks patches by the the number of passing tests. However, this category cannot rank plausible patches. The second category uses syntactic~\cite{DirectFix,dqlose,S3} and semantic distances~\cite{dqlose,S3} from the original program to rank patches. As our evaluation will show later, our approach could significantly outperform both types of distances. The third category~\cite{Long2016,Xuan2016History,xiong-icse17} learns a probabilistic model from existing rules to
rank the patches. Our approach could complement these approaches: as our evaluation will show later,
our approach is able to identify 50\% of the incorrect patches
generated by ACS, the newest approach in this category.

\smalltitle{Approaches to the Oracle Problem}
The lack of test oracle is a long-standing problem in software
testing, and the summaries of the studies on this problem can be found
in existing surveys~\cite{oracle_survey, pezze2015automated}. 
\iftoggle{longversion}{
Among
them, many studies focus on automatically generating test oracles, by
mining invariants at important programming
points~\cite{DBLP:journals/tse/ErnstCGN01,DBLP:journals/scp/ErnstPGMPTX07},
discovering important variables by mutation
analysis~\cite{DBLP:journals/tse/FraserZ12}, or using machine-learning
algorithms to classify test
runs~\cite{DBLP:conf/sigsoft/HaranKOPS05,DBLP:conf/soqua/BaahGH06}.
However, most of such approaches assume a correct
version of the program already exists, 
which does not apply
to the scenario of patch correctness identification since the
original program is already incorrect. 
On the other hand, some approaches do not require a full version of the correct program and have the potential to be applied on patch classification. 
}{
Among
them, a few studies focus on automatically generating heuristic test oracles.}
For example, invariant mining could potentially mine invariants~\cite{DBLP:journals/tse/ErnstCGN01,DBLP:journals/scp/ErnstPGMPTX07} from passing test executions to classify new test inputs. However, the effect of such an application on patch correctness identification is still unknown as far as we are aware and remains as future work.

\iftoggle{longversion}{
\smalltitle{Other related work}
Marinescu and Cadar~\cite{Marinescu:2013:KHT:2491411.2491438} propose
KATCH for testing patches. KATCH
uses symbolic execution to generate test inputs that are able to cover
the patched code. Our approach could be potentially combined with
KATCH to improve the quality of the generated tests. This is a future
direction to be explored.

Several researchers have used independent test suites~\cite{smith2015cure,Ke15ase} or using
human developers~\cite{qi15} to measure the quality of patches produced by program
repair approaches. However, these approaches cannot be used to
automatically improve the precision of existing program repair
approaches, as the independent test suites or human resources are not
available. 

Several researchers have investigated the use of the generated patches
as debugging aids. Weimer~\cite{DBLP:conf/gpce/Weimer06} has found that bug reports with generated
patches are more likely to be addressed in open source projects. Tao
et al.~\cite{Tao:2014:AGP:2635868.2635873} further found that
the quality of patches is positively correlated to debugging
correctness, where high-quality patches have a positive effect on
debugging correctness and low-quality patches have a negative effect on
debugging correctness. These studies motivate our work.
}{
\smalltitle{Other related work}
Marinescu and Cadar~\cite{Marinescu:2013:KHT:2491411.2491438} propose
KATCH for generating tests to cover patches. Our approach could be potentially combined with
KATCH to improve the quality of the generated tests. This is a future
direction to be explored.

Mutation-based fault localization such Metallaxis~\cite{Metallaxis} and MUSE~\cite{MUSE} shares a similar observation to PATCH-SIM: when mutating a faulty location, passing tests would exhibit significantly smaller behavior change than failing tests. This duality between approaches in different domains indicates that the observation is general and could potentially be applied in more domains in future.

}




\section{Patch Correctness and Behavior Similarity}\label{sec:motivation}
In this section, we analyze the relation between patch correctness and
behavior similarity to motivate our approach. We define a patch as a pair of program versions, the original buggy version and the patched version. 
To simplify discussion, we
assume the program contains only one fault. As a result, a failing test
execution must trigger the fault and produce an incorrect output.

\iftoggle{patchtype}{
\smalltitle{Types of Incorrect Patches} A patch for a fault could be incorrect in two aspects.
\begin{itemize}
\item \emph{Over-fixing}: Given an input, the original program
  produces a correct output, while the patched program produces an
  incorrect output. That is, the patch causes a regression error that
  breaks existing correct behavior.
\item \emph{Under-fixing}: Given an input, the original program
  produces an incorrect output, while the patched program still
  produces an incorrect output. That is, the patch does not repair all
  incorrect behavior.
\end{itemize}
Please note that an incorrect patch may be both over-fixing and
under-fixing at the same time\footnote{In an existing
  article~\cite{Yu17}, over-fixing and under-fixing are also known as
  A-overfitting and B-overfitting.}.

For example, \figref{fig:example1}(a) shows an incorrect patch
generated by jKali~\cite{qi15} for defect Chart-15 in the defect
benchmark Defects4J~\cite{just2014defects4j}. In this example, calling
\code{draw} will result in an undesired exception if the receiver
object is initialized with a null dataset. \figref{fig:example1}(b)
shows a test case that detects this defect by creating such an object
and checking for exceptions. \figref{fig:example1}(c) shows a passing
test checking that drawing with a null value in the dataset would not
result in an exception. This patch simply skips the whole method that
may throw the exception and is 
both under-fixing and over-fixing: the failing executions are not
properly fixed as nothing will be drawn, while the originally correct
executions are broken as the \code{draw} method is skipped.
}{}

\begin{figure}
\footnotesize
\centering
\lstset{language=Java,basicstyle=\tt\footnotesize}
\begin{lstlisting}[frame=single]
  public void draw(...) {
+   if (true) return ;
    ...
\end{lstlisting}
(a) An incorrect patch produced by jKali~\cite{qi15}
\smallskip
\begin{lstlisting}[frame=single]
public void testDrawWithNullDataset() { ...
  JFreeChart chart = ChartFactory.
    createPieChart3D("Test", null,...);
  try {...
    chart.draw(...);
    success = true; }
  catch (Exception e) {
    success = false; }
  assertTrue(success); 
}
\end{lstlisting}
(b) A failing test checking for a null dataset
\smallskip
\begin{lstlisting}[frame=single]
public void testNullValueInDataset() { ...
  dataset.setValue(..., null);
  JFreeChart chart = createPieChart3D(dataset);
  try {...
    chart.draw(...);
    success = true; }
  catch (Exception e) {
    success = false; }
  assertTrue(success); 
}
\end{lstlisting}
(c) A passing test checking for a null value in a dataset
\caption{An Incorrect Patch for Chart-15 \label{fig:example1}}
\end{figure}

\smalltitle{Weak Oracles and PATCH-SIM} As mentioned, a test suite may be weak in either  inputs or 
oracles, or both, to miss such an incorrect patch. 
\iftoggle{patchtype}{
The example in
\figref{fig:example1} demonstrates how weak oracles could miss an
incorrect patch. 
}{
To see how weak oracles could miss an incorrect patch, let us consider the example in \figref{fig:example1}. \figref{fig:example1}(a) shows an incorrect patch
generated by jKali~\cite{qi15} for defect Chart-15 in the defect
benchmark Defects4J~\cite{just2014defects4j}. In this example, calling
\code{draw} will result in an undesired exception if the receiver
object is initialized with a null dataset. \figref{fig:example1}(b)
shows a test case that detects this defect by creating such an object
and checking for exceptions. \figref{fig:example1}(c) shows a passing
test checking that drawing with a null value in the dataset would not
result in an exception. This patch simply skips the whole method that
may throw the exception. 
}
In both the passing test and the failing test, the 
oracle only checks that no exception is thrown, but does not
check whether the output of the program is correct. As a result, since
the patch prevents the exception, both tests pass.

As mentioned in the introduction, 
we rely on observation PATCH-SIM to validate patches. Given a patch, we would expect that the original
program and the patched program behave similarly on a passing test
execution, while behaving differently on a failing test execution. In this
example, the passing test would draw something on the chart in the
original program, but would skip \code{draw} method completely in the
patched program, leading to significant behavioral difference. Based on
this difference, we can determine the patch as incorrect.

\smalltitle{Weak Inputs and TEST-SIM}
To see how the weak test inputs could lead to the misses of incorrect
patches, let us consider the example in \figref{fig:example2}. This
example demonstrates an incorrect patch for Lang-39 in Defects4J
produced by Nopol~\cite{xuan2016nopol}. The original program would
throw an undesirable exception when an element in
\code{replacementList} or \code{searchList} is \code{null}. To prevent
such an exception, the correct way is to skip those elements as shown
in \figref{fig:example2}(b). However, the generated patch in
\figref{fig:example2}(a) blocks the
whole loop based on the value of \code{repeat}, which is a parameter
of the method. Interesting, all existing tests, either previously passed
or failed, happen to produce the correct outputs in the patched program. This is
because (1) the value of \code{repeat} happens to be \code{true} in
all passing tests and be \code{false} in all failing tests, and (2) the
condition \code{greater>0} is not satisfied by any element in
\code{searchList} and \code{replacementList} in the failing tests. 
\iftoggle{patchtype}{
This patch is also both over-fixing and under-fixing: it breaks correct
behavior and does not fix the defect, though neither is exposed by the
tests.}{} However, enhancing the test oracles by PATCH-SIM is not useful because the behavior on passing tests
remains almost the same as the original program while the behavior on
failing tests changes a lot.


\begin{figure}
  \centering
  \footnotesize
\lstset{language=Java,basicstyle=\tt\footnotesize,frame=single}
\begin{lstlisting}
   ...
+ if(repeat)
  for (int i = 0; i < searchList.length; i++) {
    int greater = replacementList[i].length()
      - searchList[i].length();
    if (greater > 0) increase += 3 * greater;
  }  ...
\end{lstlisting}
(a) An incorrect patch generated by Nopol~\cite{xuan2016nopol}
\lstset{language=Java}
\begin{lstlisting}
  ...
  for (int i = 0; i < searchList.length; i++) {
+   if (searchList[i] == null ||
+       replacementList[i] == null) {
+     continue;
+   }
    int greater = replacementList[i].length()
      - searchList[i].length();
    if (greater > 0) increase += 3 * greater;
  } ...
\end{lstlisting}
(b) The correct patch generated by human developers
  \caption{An Incorrect Patch for Lang-39 \label{fig:example2}}
\end{figure}


To capture those incorrect patches missed by weak test inputs, we need
new test inputs. 
To utilize PATCH-SIM with new test inputs, we need to know whether
the outputs of the tests are correct or not. To deal with this
problem, we utilize observation TEST-SIM. We assume that, when the
execution of a new test input is similar to that of a failing test, the
new test input is likely to lead to incorrect output. Similarly, when
the execution of a new test input is similar to that of a passing
test, the new test input is likely to lead to correct output. Based on
this assumption, we can classify new test inputs by comparing its execution with
those of the existing test inputs.

In the example in \figref{fig:example2}, for any new test input
triggering this bug, there will be an exception thrown in the middle of the loop,
which is similar to the executions of failing tests. On the other
hand, for any test input that does not trigger this bug, the loop will
finish normally, which is similar to the executions of existing passing
tests. 

\smalltitle{Measuring Execution Similarity} An important problem in
realizing the approach is how we measure the similarity of two test
executions. In our approach, we measure the similarity of
\emph{complete-path spectrum}~\cite{DBLP:conf/paste/HarroldRWY98} between
the two executions. A complete-path spectrum, or CPS in short, is the
sequence of executed statement IDs during a program execution. Several
existing
studies~\cite{DBLP:conf/sigsoft/DickinsonLP01,reps1997use,DBLP:conf/paste/HarroldRWY98}
show that spectra are useful in distinguishing correct and failing test
executions, and Harrold et al.~\cite{DBLP:conf/paste/HarroldRWY98}
find that CPS is among the overall best performed spectra.

For example, let us consider the two examples in \figref{fig:example1}
and \figref{fig:example2}. In both examples, the defect will lead to
an exception, which further leads to different spectra of passing and
failing tests: the passing tests will execute until the end of the method,
while the failing tests will stop in the middle. Furthermore, the failing
test executions will become different after the system is patched: no
exception will be thrown and the test executes to the end of the
method.


\smalltitle{Multiple Faults} In the case of multiple faults in a program,
the two heuristics still apply. When there are multiple faults, the
failing tests may only identify some of them. We simply treat the identified faults as one fault and the rest of them as correct program, and the above discussion still applies.


\section{Approach \label{sec:approach}}
\subsection{Overview}
\begin{figure}[!ht]
  \centering
  \includegraphics[width=0.5\textwidth]{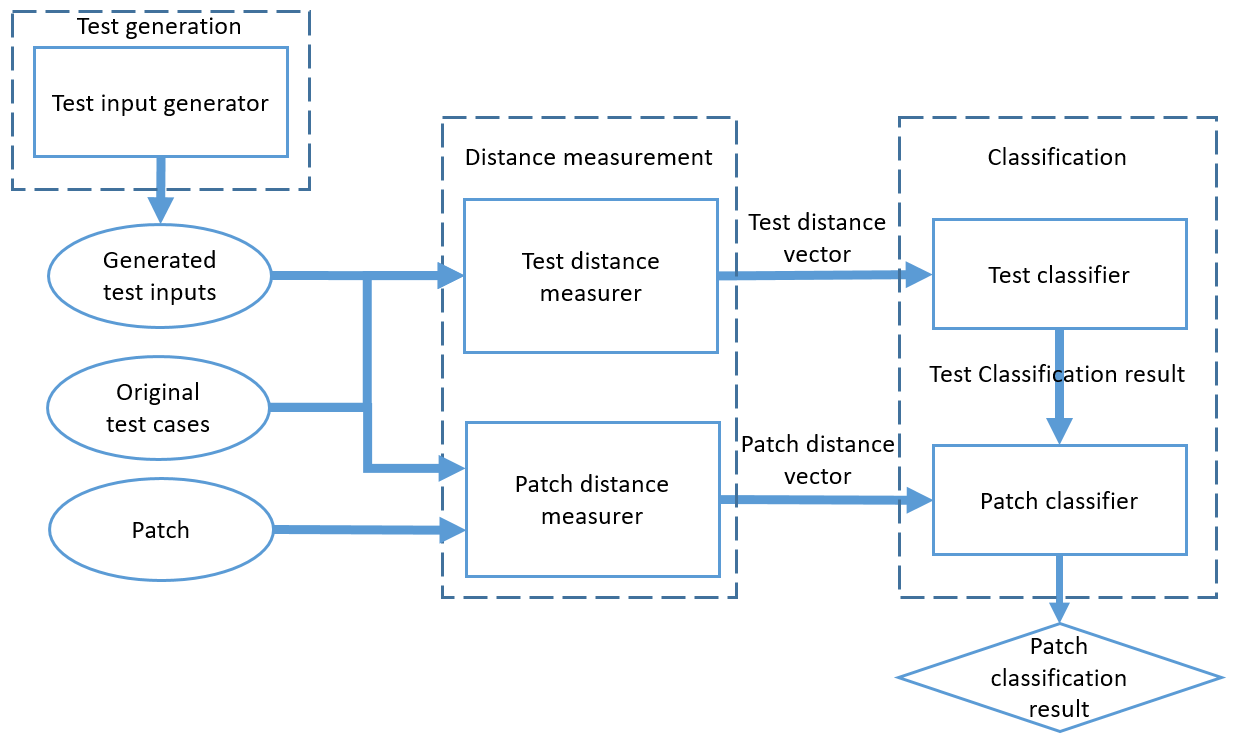}
  \caption{Approach overview \label{fig:overview}}
\end{figure}

\figref{fig:overview} shows the overall process of our approach. We
take the original buggy program, a set of test cases, and a
patch as input, and produce a classification result that tells
whether the patch is correct or not.

Our approach consists of five components classified into three
categories: test generation (including \emph{test input generator}),
distance measurement (including \emph{test distance measurer} and
\emph{patch distance measurer}) and result classification (including
\emph{test classifier} and \emph{patch classifier}). First, \emph{test
  input generator} generates a set of test inputs. We then run the
generated tests on the original buggy program. During the test
execution, we dynamically collect runtime information about the test
execution. Based on the runtime information, \emph{test distance
  measurer} calculates the distance between the executions of each
newly generated test input and each original test case. A
\emph{distance} is a real number indicating how different two test
executions are. The result is a vector of test distances. This vector
is then passed to \emph{test classifier}, which classifies the test as
{passing} or {failing} by comparing its distances to passing tests and
those to failing tests, based on TEST-SIM.


Now we have an enhanced set of test inputs which are classified as
passing or failing and we can use them to determine patch correctness.
Given a patch, \emph{patch distance measurer} runs each test on the
original program and the patched program and measure the distance
between the two executions. The result is a vector of patch distances.
Finally, this vector is taken into \emph{patch classifier} which
determines patch correctness by the distances, based on observation
PATCH-SIM.

In the rest of this section, we introduce components in the three
categories, respectively. 


\subsection{Test Generation}
Given a program, \emph{test generator} generates test inputs for this
program. Furthermore, since our goal is to determine the correctness
of the patch, we require the generated tests to cover the patched
method. If a patch modifies multiple methods, the generated tests
should cover at least one of them.

In theory, any test input generation techniques can be used in our approach.
We can utilize symbolic execution techniques to cover the specific method, especially those
designed for testing
patches~\cite{Marinescu:2013:KHT:2491411.2491438}. We can also adopt
random testing
techniques~\cite{Randoop07,fraser2011evosuite,ma2015grt} and filter
out those that do not cover any of the modified methods.

\subsection{Distance  Measurement}
\subsubsection{Measuring Distance}

As mentioned previously, we measure the distance between two
executions by comparing their complete-path spectra. As a result, the
problem reduces to measuring the distances between two sequences. In
general, there are many different metrics to measure sequence distances,
such as longest common subsequence, Levenshtein distance, Hamming
distance. As the first attempt in classifying patches based on sequence
distances, we use the longest common subsequence (LCS) as a distance
measure and leave other distance metrics to future work. 
An LCS of two sequences $a$ and $b$ is the longest sequence that can
be obtained from both $a$ and $b$ by only deleting elements. We then
normalize the length of LCS into a value between 0 and 1 using the
following formula, where $a$ and $b$ are two sequences.
\[
  distance(a,b) = 1 - \frac{|LCS(a, b)|}{max(|a|, |b|)}
\]

\subsubsection{Test Distance Measurer}
Component \emph{test distance measurer} takes a generated test and
calculates its distance with each original test. The result is a
vector of distances, where each element represents a distance between
a generated test and an original test.

To focus on the fault,
we only consider the executed statements within the calling context of
the patched methods. That is, we locate pairs of positions on the
runtime trace: (a) entering a patched method from a
method call and (b) leaving the method from the same call and keep
only the statements between the positions. This step could
help us filter noises: two tests may be different in statement
executions outside the calling context of the patched methods, but
such a difference is often not related to the fault.
If an original test does not cover any patched method, we also exclude
the test from distance measurement.

\subsubsection{Patch Distance Measurer}
Component \emph{patch distance measurer} takes each test, either
generated or original and calculates the distance between its
executions on the original program and on the patched program. The
result is a vector of distances, where each element represents the
distance of a test.

Different from \emph{test distance measurer}, here we consider the
full sequence of executed statements. This is because the compared
executions come from the same test and they are unlikely to be
noises outside the patched method.

\subsection{Classification}

Based on the distances, we can classify the generated tests and the patches. We describe the two components one by one. 
\subsubsection{Test Classifier}
The \emph{test classifier} classifies the test result of a generated
test as passing or failing. Some generated tests are difficult to
precisely classify and we discard these tests. Let $result(t)$
denotes the test result of the original test $t$, i.e., either
\emph{passing}, \emph{failing}, or \emph{discarded}. Let $distance(t, t')$ denotes the distance between the executions of $t$ and $t'$ on the original program.  
Given a generated test $t'$, we use the following formulas to determine
its classification result. The formula assigns the result of the
nearest-neighbor to the generated test. 

\begin{align*}
  &{\it classification}(t') = \left\{\begin{array}{cc}
                      \emph{passing} & \quad A_p < A_f\\
                      \emph{failing} & \quad A_p > A_f\\
                      \emph{discarded} & \quad A_p = A_f
                     \end{array}\right.\\
                   \intertext{where}
                    &A_p = min(\{distance(t, t') \mid classification(t) = passing\})\\
                   & A_f = min(\{distance(t, t') \mid classification(t) = failing\})\\
\end{align*}

Note that the above formula can only be applied when there is at least a passing test. If there is no passing test, we compare the distances with all failing tests with a threshold $K_t$ and deem the test as \emph{passing} if the test execution is significantly different from all failing tests based on the assumption that the original program works normally on most of the inputs. Please notice that there is always at least one failing test which exposes the defect.
\begin{align*}
  &{\it classification}(t) = \left\{\begin{array}{cc}
                      \emph{passing} & \quad K_t \leq A_f\\
                      \emph{failing} & \quad K_t > A_f\\
                     \end{array}\right.\\
                   \intertext{where}
                   & A_f = min(\{distance(t, t') \mid classification(t') = failing\})\\
\end{align*}




\subsubsection{Patch classifier}
\label{sec:patchclassifier}
The \emph{patch classifier} classifies a patch as \emph{correct} or
\emph{incorrect} based on the calculated
distances. Let $distance_p(t)$ denotes the distance between the
executions of test $t$ before and after applying the patch $p$. We determine the correctness of a patch $p$ using the following formula.
\begin{align*}
  &{\it classification}(p) = \left\{\begin{array}{cc}
                      \emph{incorrect} & A_p \ge K_p \\
                      \emph{incorrect} & A_p \ge A_f \\
                      \emph{correct} &  \mathrm{otherwise} \\
                     \end{array}\right.\\
\intertext{where}
  &A_p = max(\{distance_p(t) \mid classification(t) = passing\})\\
  &A_f = mean(\{distance_p(t) \mid classification(t) = failing\})\\
\end{align*}

This formula checks the two conditions in observation PATCH-SIM. First, the passing test should behave similarly. To check this condition, we compare the maximum distance on the passing tests with a threshold $K_p$ and determine the patch as incorrect if the behavior change is too large. Second, the failing test should behave differently. However, since different defects require different ways to fix, it is hard to set a fixed threshold. As a result, we check whether the average behavior change in failing tests is still larger than all the passing tests. If not, the patch is considered incorrect.

We use the maximum distance for passing tests while using the average distance for failing tests. 
An incorrect patch may affect only a few passing tests, and we use the maximum distance to focus on these tests.
On the other hand, after patched, the behaviors of all failing tests should change, so we use the average distance.

Please note this formula requires that we have at least a passing test, either original or generated. If there is no passing test, we simply treat the patch as correct.

\section{Implementation \label{sec:implementation}}
We have implemented our approach as a patch classification tool on
Java. Given a Java program with a test suite and a patch on the program,
our tool classifies the patch as correct or not.

In our implementation, we chose Randoop~\cite{Randoop07}, a random testing tool, as the test generation tool. Since our goal is to cover the patched methods, testing generation tools aiming to cover a specific location seem to be more suitable, such as the tools based on symbolic executions~\cite{Pasareanu:2010:SPS:1858996.1859035} or search-based testing~\cite{fraser2011evosuite}. However, we did not use such tools because they are designed to cover the program with fewer tests. For example, Evosuite~\cite{fraser2011evosuite} generates at most three test cases per each buggy program in our evaluation subjects. Such a small number of tests are not enough for statistical analysis.

\section{Evaluation \label{sec:eval}}

The implementation and the evaluation data are available
 online.~\footnote{\url{https://github.com/Ultimanecat/DefectRepairing}}

\subsection{Research Questions}
\begin{itemize}
\item RQ1: To what extent are TEST-SIM and PATCH-SIM reliable?
\item RQ2: How effective is our approach in identifying
  patch correctness?
\item RQ3: How is our approach compared with existing approaches,
  namely, anti-patterns, Opad, syntactic similarity and semantic similarity?
\item RQ4: How does test generation affect the overall performance?
\item RQ5: How do the parameters, $K_t$ and $K_p$, affect the overall performance?
\item RQ6: What are the causes of false positive and false negatives?
\item RQ7: How effective is our tool in classifying developers' correct patches?

\end{itemize}
RQ1 examines how much TEST-SIM and PATCH-SIM hold in general.
RQ2 focuses on the overall effectiveness of our approach. In
particular, we are concerned about how many incorrect and correct patches we filtered out. 
RQ3 compares our approach with four existing approaches for identifying
patch correctness. Anti-patterns~\cite{tananti} capture incorrect patches by matching them with pre-defined patterns. 
Opad~\cite{Yang17} is based on inherent oracles that patches should not introduce new crashes or memory-safety problem. 
Syntactic
similarity
\cite{DirectFix,dqlose,S3}
and semantic similarity~\cite{S3,dqlose} are patch ranking techniques that rank patches by measuring, syntactically or semantically, how much their changes the program, which could be adapted to determine patch correctness by setting a proper threshold. 
RQ4 and RQ5 explore how different configurations of our approach could affect the overall performance.
RQ6 investigates the causes of wrong
results in order to guide future research.
Finally, as will be seen in the next subsection, though we have tried out best to collect the generated patches on Java, the correct patches were still small in number compared with incorrect patches. To offset this, RQ7 further investigates the performance of our approach on the developers' correct patches.


\subsection{Dataset}
\begin{table*}[t]
 	 \centering
     \caption{Dataset \label{tbl:dataset}}
    \begin{tabular}{|l||l|l|l||l|l|l||l|l|l||l|l|l||l|l|l||l|l|l||l|l|l||l|l|l|}
    \hline
    \multirow{2}{*}{Project} & \multicolumn{3}{|c||}{jGenprog} & \multicolumn{3}{|c||}{jKali}    & \multicolumn{3}{|c||}{Nopol2015} & \multicolumn{3}{|c||}{Nopol2017} & \multicolumn{3}{|c||}{ACS}      &\multicolumn{3}{|c||}{HDRepair}	& \multicolumn{3}{|c||}{Total(Generated)}  & \multicolumn{3}{|c||}{Developer Patches}\\ 
    \cline{2-25}
     & P & C & I & P & C & I & P & C & I & P & C & I & P & C & I & P & C & I & P & C & I & P & C & I\\
     \hline
    Chart   & 6 & 0 &6   & 6 & 0 &6  & 6 &1 & 5    & 6 &0 &6    & 2 &2 &0   &0 &0 &0	& 26&3&23          & 25&25&0 \\
    Lang    & 0 & 0 &0        & 0 &0 &0        & 7&3&4    & 4&0&4    & 3&1&2   &1&0&1	& 15&4&11          & 58&58&0 \\
    Math    &  14&5&9 & 10 &1 &9  & 15&1&14  & 22&0&22  & 15&11&4 &7&2&5	& 83&20&63         & 84&84&0 \\
    Time    & 2&0&2   & 2&0 &2   & 1&0&1    & 8&0&8    & 1&1&0   &1&1&0	& 15&2&13          & 27&27&0 \\ 
    \hline
    Total   & 22&5&17 & 18 &1 &17 & 29&5&24  & 40&0&40  & 21&15&6 &9&3&6	& 139&29&110       & 194&194&0\\ \hline
    \end{tabular}\\
P=Patches, C=Correct Patches, I=Incorrect Patches
\end{table*}


We have collected a dataset of generated patches from existing papers. 
\tblref{tbl:dataset} shows the statistics of the dataset. Our dataset
consists of patches generated by six program repair tools. Among the
tools, jGenProg is a reimplementation of GenProg~\cite{GenProg,GenProgTSE,EightDollar} on Java, a repair tool based on
genetic algorithm; jKali is a reimplementation of Kali~\cite{qi15} on Java, a
repair tool that only deletes functionalities; Nopol~\cite{xuan2016nopol} is a tool that relies
on constraint solving to fix incorrect conditions and two versions,
2015~\cite{martinez2015automatic} and 2017~\cite{xuan2016nopol}, are used in our experiment; HDRepair~\cite{Xuan2016History} uses information from historical bug fixes to guide the search process; ACS~\cite{xiong-icse17} is a tool based on
multiple information sources to statistically and heuristically fix
incorrect conditions. 
The selected tools cover the three types
of patch generation approaches: search algorithms (jGenProg, jKali),
constraint-solving (Nopol) and statistical (HDRepair, ACS). More details of the three types can be found in the related work section. 

The
patches generated by jGenProg, jKali and Nopol2015 are collected from
Martinez et al.'s experiments on
Defects4J~\cite{martinez2015automatic}. The patches generated by
Nopol2017 are collected from a recent report on Nopol~\cite{xuan2016nopol}.
Patches generated by HDRepair is obtained from Xin and Reiss' experiment on patch classification~\cite{Xin17}.
The patches generated by ACS is collected from ACS
evaluation~\cite{xiong-icse17}.

All the patches are generated for defects in Defects4J~\cite{just2014defects4j},
a widely-used benchmark of real defects on Java. Defects4J consists of
six projects: Chart is a library for displaying charts; Math is a
library for scientific computation; Time is a library for date/time
processing; Lang is a set of extra methods for manipulating JDK classes;
Closure is optimized compiler for Javascript; Mockito is a mocking
framework for unit tests. 

Some of the patches are not supported by our implementation, mainly because Randoop cannot generate any tests for these patches. In particular, Randoop cannot generate any tests for Closure and Mockito. We removed these unsupported patches.

The patches from Martinez et al.'s experiments, the ACS evaluation and Qi et al.'s experiments contains labels identifying the correctness of the patches, which mark the patches as \emph{correct}, \emph{incorrect}, or \emph{unknown}. The
patches of Nopol2017 do not contain such labels. We manually checked whether 
the unlabeled patches and some labeled patches are semantically equivalent to the human-written patches. Since the patches whose correctness is
unknown cannot be used to evaluate our approach, we remove these
patches. 


In the end, we have a dataset of 139 patches generated by automatic program repair tools, where 110 are
incorrect patches and 29 are correct patches. 

To answer RQ6, we also added all developer patches on Defects4J into our dataset.  Same as generated patches, we removed the unsupported patches, including all patches on Closure and Mockito. In the end, we have 194 developer patches. Please note that developer patches are only used in RQ6 since they have different characteristics compared with generated patches.


\subsection{Experiment Setup}


\smalltitle{Test Generation}  We kept Randoop to run 3 minutes on the original program and collected the tests that covered the patched methods. We stop at 3 minutes because for most defects, Randoop produced enough tests within three minutes, and for the remaining defects that do not have enough tests, lengthening the time would not lead to more tests. We then randomly selected 20 tests for each patch. If there were fewer than 20 tests, we selected all of them. In the end, we have 7.1 tests per patch in average, with a minimum of 0 test. Based on the classification of TEST-SIM, 71\% of the generated tests are passing tests. 

\smalltitle{RQ1}
To evaluate PATCH-SIM, we measured the average distance between test executions of patched and unpatched versions, and check whether there is significant differences between passing and failing tests on correct and incorrect patches. 
To evaluate TEST-SIM, we measured the distances between tests, and analyzed whether closer distances indicate similar test results. 

\smalltitle{RQ2}
We applied our approach to the patches in our dataset and
checked whether our classification results are consistent with the labels
about correctness. 

\smalltitle{RQ3}
We applied the four existing approaches to the dataset
and compared their results with our result. 

Anti-patterns was originally implemented in C. To apply anti-patterns on Java dataset, we
took the seven anti-patterns defined by Tan et al.~\cite{tananti} and
manually checked whether the patches in our dataset fall into these
patterns.

Opad~\cite{Yang17} uses inherent oracles that patches should not introduce new crash or memory-safety problems. Opad was originally designed for C and we need to adapt it for Java. On the one hand, crashes are represented as runtime exceptions in Java. On the other hand, memory-safety problems are either prevented by the Java infrastructure or detected as runtime exceptions. Therefore, we uniformly detect whether a patch introduces any new runtime exception on test runs. If so, the patch is considered incorrect.


Regarding syntactic and semantic distances, different papers~\cite{DirectFix,dqlose,S3} have proposed different metrics to measure the syntactic and semantic distances and a summary can be found in \tblref{tbl:metrics}. However, as analyzed in the table, many of the metrics are defined for a specific category of patches and cannot be applied to general patches. In particular, many metrics are designed for expression replacement only and their definitions on other types of changes are not clear.
As a result, we chose the two metrics marked as "general" in \tblref{tbl:metrics} for comparison: one measuring syntactic distance by comparing AST and one measuring semantic distance by comparing complete-path spectrum.



\begin{table*}[ht]
	\centering
    \caption{Different syntactic and semantic metrics \label{tbl:metrics}}
    \begin{tabular}{|p{3cm}|c|p{1.8cm}|p{11cm}|}
    \hline
    Metric  & Type  & Scope & Description\\
    \hline
    AST-based \cite{DirectFix,S3}  & Syn  & General & Number of AST node changes introduced by the patch.\\
    Cosine similarity \cite{S3} & Syn 	&  Replacement  & The cosine similarity between the vectors representing AST node occurrences before and after the change. It is not clear how to apply it to insertions and deletions because there will be a zero vector and cosine similarity cannot be calculated.   \\
    Locality of variables and constants \cite{S3} & Syn  & Expression Replacement & The distance is measured by the Hamming distance between the vectors representing locations of variables and constants. It is not clear how to apply it to patches with multiple changes. \\
    Expression size distance \cite{dqlose} & Syn &  Expression Replacement & The distance is 0 when two expressions are identical, otherwise the size of the affected expression after patching. It is not clear how to apply it to changes other than expression replacement.\\
    Complete-path spectrum \cite{dqlose} & Sem  & General & The difference between complete-path spectra.           \\
    Model counting \cite{S3} & Sem &  Boolean Expression replacement  & The distance is measured by the number of models that make two expressions evaluate differently.  The definition is bounded to Boolean expressions. \\
    Output coverage \cite{S3} & Sem  & Programs with simple outputs  & The distance is measured by the proportion of different outputs covered by the patched program. It is not clear how to define ``output'' in general for complex programs. \\
    \hline
    \end{tabular}
    "Syn" and "Sem" stand for syntactical distance and semantic distance respectively.
\end{table*}

The AST-based syntactic distance is defined as the minimal number of AST nodes that need to be deleted or inserted to change the one program into the other program. For example, changing expression $a > b + 1$ to $c < d + 1$ needs to at least remove three AST nodes ($a$, $b$, $>$) and insert three AST nodes ($c$, $d$, $<$), giving a syntactic distance of 6. The semantic distance based on complete-path spectrum for program $p$ is defined using the following formula, where  $T_o$ is the set of all original tests that cover at least one modified method and $distance_p$ is defined in   \secref{sec:patchclassifier}.
\begin{align*}
  &LED(p) = mean(\{distance_p(t) \mid t \in T_o \})
\end{align*}

The syntactic/semantic distance gave a ranked list of the patches. Then we checked if we could find an optimal threshold to separate the list into correct and incorrect patches.


\smalltitle{RQ4}
We considered two different test generation strategies and
compared their results with the result of RQ1.
\begin{itemize}
\item {\bf No generation.} This strategy simply does not generate any test
  input. This strategy serves as a baseline for evaluating how much
  the newly generated test inputs contribute to the overall performance.
\item {\bf Repeated Randoop runs.} Since Randoop is a random test
  generation tool, different invocations to Randoop may lead to
  different test suites. This strategy simply re-invokes Randoop to
  generate a potentially different set of tests and the comparison
  helps us understand the effect of the randomness in test generation
  on the overall performance of our approach.

\end{itemize}
Since the second strategy involves re-generating the tests and is expensive to perform, we evaluated this strategy on a randomly selected subset of 50 patches. To see the effect of randomness, we repeated the experiments for 5 times.

\smalltitle{RQ5}
During the experiments for the first three research questions, we set the parameters of our approach as follows: $K_p$ = 0.25, $K_t$ = 0.4. These parameter values are determined by a few attempts on a small set of patches. 

To answer RQ4, we systematically set different values for parameters $K_p$ and $K_t$ and then analyzed how these parameters affect our result on the whole dataset.


\smalltitle{RQ6}
We manually analyzed all false positives and false
negatives to understand the causes of false classification and
summarize the reasons.

\smalltitle{RQ7}
We applied our approach on human-written patches provided by Defects4J benchmark and check whether our approach misclassified them as incorrect or not.

\smalltitle{Hardware Platform}
The experiment is performed on a server with Intel Xeon E3 CPU and 32GB memory. 

\subsection{Result of RQ1: Reliability of Heuristics}

\begin{table}[!ht]
	\centering
    \caption{PATCH-SIM \label{tbl8:patch-sim}}
    \begin{tabular}{|l|ll|}
    \hline
    ~  & Passing Tests  & Failing Tests \\
    \hline
    Incorrect Patches & 0.25       & 0.33 \\
    Correct Patches & 0.02        & 0.19     \\

    \hline
    \end{tabular}
\end{table}

Table~\ref{tbl8:patch-sim} shows the results of evaluating PATCH-SIM, i.e., the distances between test executions on patched and unpatched versions. As we can see from the table, for correct patches, the distances of passing tests are very close to zero, while failing tests have a much larger distances that is 9.5 times of passing tests. The result indicates that PATCH-SIM holds in general. On the other hand, the passing tests and the failing tests do not exhibit such a strong property on incorrect patches. While the distances of failing tests are still larger than passing tests, the ratio is only 1.32 times rather than 9.5 times. This results indicate that PATCH-SIM can be used to distinguish correct and incorrect patches.

\begin{figure}[!ht]
  \centering
  \includegraphics[width=\columnwidth]{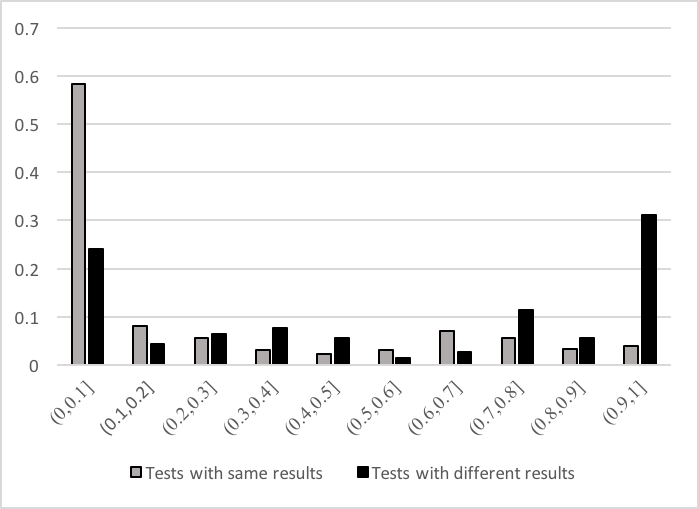}
  \parbox{0.8\columnwidth}{\footnotesize X-axis: intervals of distance on tests \ \ \ \ Y-axis: percent of tests}
    \caption{TEST-SIM \label{fig:test-sim}}
\end{figure}

Figure~\ref{fig:test-sim} shows the results of evaluating TEST-SIM. The X-axis shows intervals of distances while the Y-axis shows the percentage of tests fall into the intervals. As we can see from the figure, when two tests have a short distance, they are more likely to have the same test results rather than different test results. This result indicates that TEST-SIM holds in general. On the other hand, when the two tests have a long distance, they are more likely to have different test results rather than the same test results. 

\subsection{Result of RQ2: Overall Effectiveness}

\begin{table}[!ht]
    \centering
    \caption{Overall Effectiveness per Tool \label{tbl2:overview}}
    \begin{tabular}{|c|ccp{1.3cm}p{1.3cm}|}
    \hline
    Tool      & Incorrect & Correct & Incorrect Excluded &  Correct Excluded\\ \hline
    jGenprog  & 17            & 5               & 8(47.1\%)                          & 0                        \\
    jKali     & 17            & 1               & 9(52.9\%)                          & 0                        \\
    Nopol2015 & 24            & 5               & 16(66.7\%)                        & 0                        \\
    Nopol2017 & 40            & 0               & 22(55.0\%)                       & 0                        \\
    ACS       & 6             & 15              & 3(50.0\%)                        & 0                        \\
    HDRepair       & 6             & 3              & 4(66.7\%)                        & 0                       \\
    \hline
    Total     & 110          & 29               & 62(56.3\%)   &   0\\
    \hline
    \end{tabular}
    
    ``In/correct Excluded'' shows the number of patches that are filtered out by our approach and are in/correct. 
    
\end{table}

\begin{table}[!ht]
    \centering
    \caption{Overall Effectiveness per Project \label{tbl3:projects}}
    \begin{tabular}{|c|ccp{1.3cm}p{1.3cm}|}
    \hline
    Project      & Incorrect & Correct & Incorrect Excluded & Correct Excluded \\ \hline
    Chart     & 23            & 3               & 14(60.9\%)                          & 0                        \\
    Lang & 11            & 4              & 6(54.5\%)                       & 0                        \\
    Math & 63            & 20               & 33(52.4\%)                        & 0                        \\
	Time  & 13            & 2               & 9(69.2\%)                          & 0                        \\
\hline
Total     & 110          & 29               & 62(56.3\%)   &   0\\
    \hline
    \end{tabular}
\end{table}

\tblref{tbl2:overview} and \tblref{tbl3:projects} shows the performance of our approach on the dataset per tool and per project, respectively. 
As shown in the tables, our approach successfully filtered out 62 of 110 incorrect plausible patches and filtered out no correct patch. Furthermore, our approach shows similar performance on different tools and different projects, indicating that our results are potentially generalizable to different types of projects and different types of tools.

Please note that although our approach did not filter out any correct patch on our dataset, in theory it is still possible to filter out correct patches. For example, a patch may significantly change the control flow of a passing test execution, e.g., by using a new algorithm or calling a set of different APIs, but the test execution could produce the same result. However, given the status of current program repair approaches, such patches are probably scarce.
When applying our approach on human-written patches, some correct patches are filtered out. More details for the effectiveness on human-written patch is discussed in RQ6.



Our approach took about 5 to 10 minutes to determine the correctness of a patch in most cases, while some patches might take up to 30 minutes. Most of the time was spent on generating the test inputs and recording the runtime trace.

\subsection{Result of RQ3: Comparing with Others}

\smalltitle{Anti-patterns}
Among all 139 patches, anti-patterns filtered out 28 patches, where 27 are incorrect and 1 is correct. The result shows that our approach significantly outperforms anti-patterns. Furthermore, 13 of the 27 incorrect patched filtered out by anti-patterns were also filtered out by our approach, while the remaining 14 patches were not filtered out by our approach. This result suggests that we may potentially combine the two approaches to achieve a better performance.

\smalltitle{Opad}
When applied with the same set of test inputs as our approach, Opad failed to recognize any of the incorrect patches. To further understand whether a stronger test suite could achieve better results, we further selected up to 50 tests instead of 20 tests for each patch. This time Opad filtered out 3 incorrect patches. This result suggests that inherent oracles may have a limited effect on classifying Java patches, as the Java infrastructure has already prevented a lot of crashes and memory safety problems and it may not be very easy for a patch to break such an oracle.


\begin{figure}[!ht]
  \centering
  \includegraphics[width=\columnwidth]{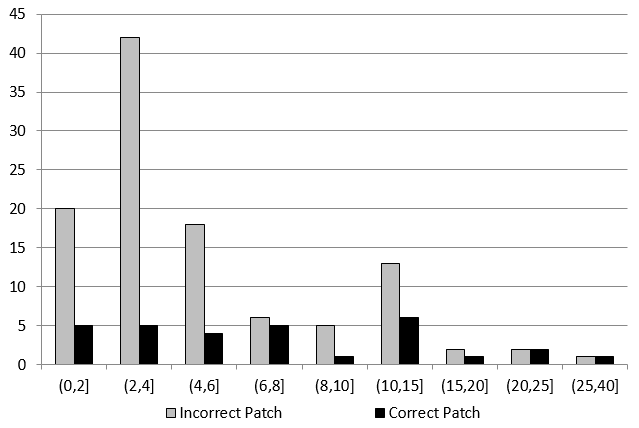}
  \parbox{0.8\columnwidth}{\footnotesize X-axis: intervals of syntactic distance \ \ \ \  Y-axis: numbers of patches}
  \caption{Syntactic distance \label{fig:syn-dis}}
\end{figure}

\smalltitle{Syntactic and Semantic Distance}
Fig. \ref{fig:syn-dis} shows the distribution of incorrect patches and correct patches on syntactic distance. The x-axis shows the intervals of distances while Y-axis shows the numbers of patches within the intervals. As we can see from the figure, the incorrect patches and correct patches appear in all intervals and the distribution shows no particular characteristics. If we would like to exclude 56.3\% incorrect patches using syntactic distance, we need to at least exclude 66.7\% of the correct patches. This result indicates that syntactic distance cannot be directly adapted to determine patch correctness.

\begin{figure}[!ht]
  \centering
  \includegraphics[width=\columnwidth]{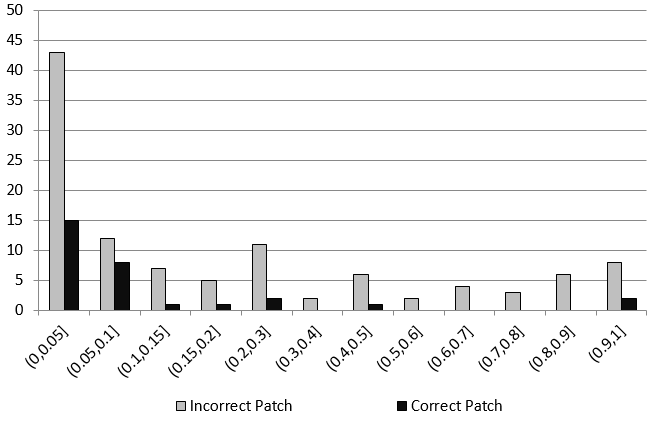}
  \parbox{0.8\columnwidth}{\footnotesize X-axis: intervals of semantic distance \ \ \ \ Y-axis: numbers of patches}
    \caption{Semantic distance \label{fig:sem-dis}}
\end{figure}

Fig. \ref{fig:sem-dis} shows the distribution of incorrect and correct patches on semantic distance. As we can see from the figure, both types of patches tend to appear more frequently when the distance is small. When the distance grows larger, both types decrease but correct patches decrease faster. If we would like to exclude 56.3\% incorrect patches using semantic distance, we need to exclude 43.3\% correct patches. This result indicates that semantic distance could be a better measurement than syntactic distance in determining patch correctness, but is still significantly outperformed by our approach.


Please note that the above results do not imply that syntactic and semantic distances are not good at ranking patches. While it is difficult to find a threshold to distinguish correct and incorrect patches for a group of defects, it is still possible that the correct patches are ranked higher on most individual defects.

\subsection{Result of RQ4: Effects of Test Generation}
\tblref{tbl3:test generation} shows the result without generating test inputs. Without the generated test inputs, our approach filtered out 8 less incorrect patches and still filtered 0 correct patch. This result suggests that PATCH-SIM alone already makes an effective approach, but test generation and TEST-SIM can further boost the performance non-trivially. 
\begin{table}[!ht]
	\centering
    \caption{Comparison with no test generation \label{tbl3:test generation}}
    \begin{tabular}{|l|ll|}
    \hline
    ~  & Default Approach  & No Generation \\
    \hline
    Incorrect Excluded & 62       & 54 \\
    Correct Excluded & 0        & 0     \\
    \hline
    \end{tabular}
\end{table}




Regarding randomness, we repeated our experiment on the selected 50 patches 5 times and got different results on only 3 incorrect patches. The best case had only one more excluded incorrect patch than the worst case. The result shows that randomness does affect the result, but the impact is limited.


\subsection{Result of RQ5: Parameters}
Two parameters are involved in our approach, $K_t$ and $K_p$, both ranging from 0 to 1. 
The results of our approach with different parameters are shown in
Table \ref{tbl4:parameterKp} and \ref{tbl5:parameterKt}, where each column shows the result with the parameter value in the
table head. As we can see from the tables, setting the parameters to
different values have a limited impact on the overall results and a
large range of parameter value could achieve the best performance. The
result indicates that our approach does not require a precise tuning
of parameters.

\begin{table}
	\centering
     \caption{Parameter $K_p$ \label{tbl4:parameterKp}}
    \begin{tabular}{|l|llllllllll|}
    \hline
    ~  & 0.05 & 0.1 & 0.15 & 0.25 & 0.3 & 0.4 & 0.5 & 0.6 & 0.8 & 1 \\ \hline
    IE & 71   & 66  & 62   & 62   & 60  & 57  & 56  & 56  & 55  & 54 \\
    CE & 4    & 1   & 0    & 0    & 0   & 0   & 0   & 0   & 0   & 0 \\ \hline
    \end{tabular}
    IE = Incorrect Excluded, CE = Correct Excluded
\end{table}

\begin{table}
	\centering
     \caption{Parameter $K_t$ \label{tbl5:parameterKt}}
    \begin{tabular}{|l|lllllllllll|}
    \hline
    ~  & 0  & 0.1 & 0.2 & 0.3 & 0.4 & 0.5 & 0.6 & 0.7 & 0.8 & 0.9 & 1  \\ \hline
    IE & 65 & 62  & 62  & 62  & 62  & 60  & 60  & 60  & 60  & 59  & 59 \\
    CE & 4  & 1   & 0   & 0   & 0   & 0   & 0   & 0   & 0   & 0   & 0  \\ \hline
    \end{tabular}
    IE = Incorrect Excluded, CE = Correct Excluded
\end{table}

\subsection{Result of RQ6: Causes of Wrong Result}

Our approach gave wrong results on 47 incorrect
patches. We manually analyzed these patches and identified three main causes of the wrong classification, as follows.

\smalltitle{Too weak test suite}
It is often the case (21 out of 48) that only one failing test covers the patched method. Without passing test, our approach only relies on the threshold $K_t$  to classify the tests and it is sometimes difficult to generate tests that pass the threshold. As a result, we might have no or only a few passing tests to perform the patch classification, leading to a low performance.



\smalltitle{Unsatisfying test generation}
Another common case (27 out of 48, 9 overlap with the previous case) is that test generation tool fails to generate satisfying tests. Randoop might fail to give tests that cover the patched method or  fail to generate tests that could expose the incorrect behavior. The patches in this category have the potential to be correctly identified if we use a stronger test generation tool.

\smalltitle{Unsatisfying classification formula}
The final case (8 out of 48) was caused by large behavior changes in some failing test execution. Since we calculated the average distance of all failing test executions in the patch classification, if there was a very large value, the average value might become large even if all the rest failing tests had small behavior changes. As a result, the patch may be misclassified.
This problem may be fixed by generating more failing test cases to lower down the average value, or to find a better formula to classify the patches. 



\subsection{Result of RQ7: Developer Patches}

Among the 194 correct developer patches, our tool classified 16 (8.25\%) patches as incorrect.  We further analyzed why the 16 patches are misclassified and found that all the 16 patches have non-trivially changed the control flow and caused a significant difference in CPS in the passing test executions. In particular, the behaviors of passing tests have significantly changed in 6 patches, while in the rest 10 patches the behaviors remain almost the same but the executed statements changed significantly (e.g., calling a different method with the same functionality).  The results imply that (1) human patches are indeed more complex than those generated by current automated techniques; (2) when the complexity of patches grows, our approach is probably still effective as only a small portion of correct patches is excluded; (3) To further enhance the performance, we need to enhance PATCH-SIM and CPS to deal with such situations. 



\section{Threats to Validity and Limitations \label{sec:threats}}
The main threat to internal validity is that we discarded some patches
from our dataset, either because their correctness cannot be
determined, or because the infrastructure tool used in our
implementation cannot support these patches. As a result, a selection
bias may be introduced. However, we
believe this threat is not serious because the removed patches are small in number compared with the whole dataset and the results on these patches are unlikely to significantly change the overall results.


The main threat to external validity is whether our approach can be
generalized to different types of program repair tools. While we have selected repair tools from all main categories of
program repair tools, including tools based on search algorithms,
constraint solving and statistics, it is still unknown
whether future tools will have characteristics significantly different
from current tools. To minimize such a threat, we have added RQ7 to test on developer patches, which can be viewed as the ultimate goal of automatically generated patches. The results indicates that our approach may have different performance on developer patches and generated patches, but the difference is limited. 

The main threat to construct validity is that the correctness of the
patches are manually evaluated and the classification may be wrong.
To reduce this threat, all difficult patches are discussed through the
first two authors to make a mutual decision. Furthermore, part of the
classification comes from Martinez et al.'s
experiment~\cite{martinez2015automatic}, whose results have been
published online for a few years and there is no report questioning
the classification quality as far as we are aware.

There can be many different choices in designing the formulas. For
example, we can use a different sequence distance or even a different
spectrum to measure the distance of two executions. We can use
different statistical methods for classifying tests and patches. The
current paper does not and cannot explore all possibilities and leave
them as future work.


 {}

\section{Conclusion \label{sec:conclusion}}
In this paper, we have proposed an approach to automatically determining
the correctness of the patches based on behavior similarities between
program executions. As our evaluation shows, our approach could
effectively filter out 56.3\% of the incorrect patches generated without losing any of the correct patches.
The result suggests that measuring behavior similarity can be a promising way to
tackle the oracle problem and calls for more research on this topic.



%
\bibliographystyle{ACM-Reference-Format}
\bibliography{references}

\end{document}